\documentstyle[11pt,bezier,epsf]{article}

\newcommand{\sect}[1]{ \section{#1} \setcounter{equation}{0} }

\newcommand{\half}{\mbox{\small{$\frac{1}{2}$}}}

\newcommand{\MSbar}{\overline{\mbox{MS}}}

\begin{document}
\hfill SPbU--IP--97--07

\hfill LTH--391

\begin{center}
{\LARGE {Four loop anomalous dimensions of gradient operators in $\phi^4$
theory}} \\ [8mm]
S.\'{E}. Derkachov$^a$\footnote{e-mail: derk@tu.spb.ru},
J.A. Gracey$^b$\footnote{e-mail: jag@amtp.liv.ac.uk}
\& A.N. Manashov$^c$\footnote{e-mail: manashov@snoopy.phys.spbu.ru} \\ [3mm]
\begin{itemize}
\item[$^a$] Department of Mathematics, St Petersburg Technology Institute, \\
Sankt Petersburg, Russia.
\item[$^b$] Theoretical Physics Division, Department of Mathematical Sciences,
\\
University of Liverpool, Liverpool, L69 7ZF, United Kingdom.
\item[$^c$] Department of Theoretical Physics, State University of St
Petersburg, \\ Sankt Petersburg, 198904 Russia.
\end{itemize}
\end{center}

\vspace{3cm}
\begin{abstract}
We compute the anomalous dimensions of a set of composite operators which
involve derivatives at four loops in $\MSbar$ in $\phi^4$ theory as a function
of the operator moment $n$. These operators are similar to the twist-$2$
operators which arise in QCD in the operator product expansion in deep
inelastic scattering. By regarding their inverse Mellin transform as being
equivalent to the DGLAP splitting functions we explore to what extent taking a
restricted set of operator moments can give a good approximation to the {\em
exact} four loop result.
\end{abstract}

\newpage

\sect{Introduction.}

The scalar field theory with a $\phi^4$ interaction has been widely studied
for a variety of problems. For instance, it underlies the physics of the
anti-ferromagnetic phase transition in statistical physics and also is the
starting point for the Higgs mechanism of the standard model in particle
physics. From another viewpoint it has been used as a toy model in four
dimensions to examine fundamental ideas in quantum field theory. One such
example is understanding perturbation theory at high orders. (For a review of
these points see, for example, \cite{jzj}.) In particular the fundamental
functions of the renormalization group like the $\beta$-function are known to
five loops in the $\MSbar$ scheme, \cite{kub,kle}. In other field theories of
interest in particle physics such as gauge theories four loop results have only
become available in the last few years. For instance the four loop
$\beta$-functions of QED and QCD, which describes the strong interactions, were
computed in \cite{qed4,qcd4} respectively.

Such calculations are necessary in QCD primarily because it is asymptotically
free and therefore it describes the physics of high energy collisions
involving quarks. As these deep inelastic experiments are becoming
increasingly more accurate the theoretical input must be refined accordingly.
Therefore current perturbative calculations in QCD have been focussed on the
contribution of the anomalous dimensions and coefficient functions of the
twist-$2$ flavour non-singlet and singlet operators that arise in the operator
product expansion. These were originally computed at one and two loops in
\cite{gro,cts} as a function of the moment of the operator $n$. Recently the
first few moments of the three loop dimensions were determined in
\cite{nsdis,sdis}. However, the programme is not yet complete as one requires
the explicit $n$-dependence at three loops. This is required in order to
compute the DGLAP splitting functions, \cite{dglap}, which are given by the
inverse Mellin transform with respect to $n$. In \cite{sdis} an approximate
fitting procedure was developed to obtain these functions. This was achieved by
using general information on the (expected) form of the function in Bjorken
$x$, which is the variable conjugate to $n$ with $0$ $\leq$ $x$ $\leq$ $1$.
Also the information from the explicit exact result for the first few moments
was used to constrain the fitting parameters. In the absence of a full
$n$-dependent result for the operator dimensions this strategy has to suffice
for present. Clearly there are several potential problems in such an exercise.
For example, the first few moments would only be expected to give a fairly
accurate approximation to the splitting functions in the bulk of the $x$-range.
However, the low $x$ region which is presently of physical interest would not
be well covered since there the higher moments give important contributions. So
it would be useful to understand to what extent such a fitting procedure can be
relied upon.

Whilst the complete calculation in QCD will be a formidable computational task
one could at least address this and other issues in a toy model. This is the
primary motivation for this paper. We will compute the anomalous dimensions of
the analogous operators in $\phi^4$ theory to four loops in the coupling
constant expansion in the $\MSbar$ scheme. Although this would appear to be one
order beyond that currently of interest in QCD, it turns out that due to the
form of the model there can be no one loop contribution. We view this theory as
a toy for several reasons. First, the nature of the interaction is simpler
than in a non-abelian gauge theory and hence there will be substantially fewer
diagrams to consider and also no tensorial complication in the numerator of the
Feynman diagrams which occurs when one has fermion and derivative couplings.
Consequently although we do not expect the results to be as involved as those
which will undoubtably occur when the full QCD result is realised, those
functions of $n$, such as finite sums in $n$, which will occur will at least
mimic analogous structures. Indeed as will be seen, it is these forms which
will drive the strategy of the fitting procedure. As we will obtain the full
$4$-loop results as a function of $n$ we will be in a position to investigate
and gain insight into the approximation procedure for the splitting functions.

This leads us to our second reason why we will regard our work as a laboratory
for testing ideas. Clearly unlike QCD $\phi^4$ theory in four spacetime
dimensions is not asymptotically free and therefore it does not sense to think
of the high energy limit describing particle collisions. However we stress that
the aim is a {\em mathematical} one to endeavour to understand the relation of
functions of the operator moment $n$ to their Mellin transform which in
$\phi^4$ we will regard as being on the same footing as the DGLAP functions. In
this situation asymptotic freedom plays a minimal role. Finally, our second
motivation for this study is in relation to understanding the content of the
operator product expansion of $\phi^4$ in $d$-dimensions. As noted earlier the
model relates to the perturbation theory of the Heisenberg model which
equivalently, in the sense of critical phenomena, can be described by the
$O(N)$ non-linear $\sigma$ model
in $d$ $=$ $(2$ $+$ $\epsilon)$ dimensions. (See,
for example, \cite{jzj}.) The operator product expansion in that model has
received wide attention in recent years both perturbatively and in the $1/N$
expansion, \cite{rlwk}. Therefore the computation of the anomalous dimension of
a set of operators involving derivatives or gradients at high order in the
related $\phi^4$ theory is important for providing at least a cross check on
future perturbative calculations in this area.

The paper is organised as follows. In section $2$, we review basic features of
the renormalization of composite operators in quantum field theories and then
perform the two and three loop computations for the anomalous dimensions of the
operators in question. This calculation is extended to the fourth order in
section $3$ where we also discuss the technical details of the determination of
the divergent part of some of the underlying Feynman integrals. These results
are then used to deduce the corresponding splitting functions in section $4$
where the approximate fitting procedure is analysed. Our conclusions are given
in section $5$ and intermediate results in the four loop calculations are given
in an appendix.

\sect{Preliminaries.}

We begin our discussion by reviewing the background to the computation of
anomalous dimensions of composite operators. We will concentrate on one
particular set of operators in this paper,
\begin{equation}
{\cal O}_n ~=~ \phi(x) \partial_{\mu_1} \ldots \partial_{\mu_n} \phi (x)
{}~-~ \mbox{traces}
\end{equation}
which are twist-$2$ from the phenomenology point of view and symmetric and
traceless. The field $\phi$ is the basic field of the $\phi^4$ theory which
has the lagrangian
\begin{equation}
L ~=~ \frac{1}{2} (\partial \phi)^2 ~-~ \frac{g}{24} (\phi^2)^2
\end{equation}
where $g$ is the coupling constant which is dimensionless in
$4$-dimensions. As we will be applying standard techniques of renormalization
such as dimensional regularization in $d$ $=$ $4$ $-$ $2\epsilon$ dimensions
and using the modified minimal subtraction scheme, ($\MSbar$), we must derive
the momentum space version of the operator. Its divergent structure is
deduced from the divergence structure of $2$-point Green's functions where
${\cal O}_n$ has been inserted. Therefore we write
\begin{equation}
\int \frac{d^dx}{(2\pi)^d} \, {\cal O}_n ~=~
\int \frac{d^dp}{(2\pi)^d} \, \tilde{\phi}(-p) T_n(p)_{\mu_1 \ldots \mu_n}
\tilde{\phi}(p)
\end{equation}
where $T_n(p)_{\mu_1 \ldots \mu_n}$ is a symmetric traceless tensor. It is
more convenient to swamp the Lorentz indices by contracting with a constant
vector $\Delta_\mu$ and define the object
\begin{equation}
T_n(p,\Delta) ~=~ T_n(p)_{\mu_1 \ldots \mu_n} \Delta^{\mu_1} \ldots
\Delta^{\mu_n}
\end{equation}
Given the fact that $T_n(p)$ is traceless one can derive the form of
$T_n(p,\Delta)$. However, in higher order calculations we will exploit the
first two terms in the construction of recurrence relations for subintegrals.
Therefore we record,
\begin{equation}
T_n(p,\Delta) ~=~ (\Delta p)^n ~-~ \frac{n(n-1)}{4(n+d-3)} p^2 \Delta^2
(\Delta p)^{n-2} ~+~ O( (\Delta^2)^2 (\Delta p)^{n-4} )
\label{Tdefn}
\end{equation}
In QCD calculations \cite{gro,cts,nsdis,sdis} the constant vector is usually
taken to be a null vector in which case (\ref{Tdefn}) terminates at the first
term.

The anomalous dimension we are aiming to compute is $\gamma_n(g)$. In terms
of renormalization constants this is defined to be
\begin{equation}
\gamma_n(g) ~=~ M^2 \frac{\partial~~}{\partial M^2}
\ln (Z_{1}^{-1} Z_{\cal O})
\end{equation}
where $M$ is the mass scale introduced to ensure $g$ remains dimensionless in
$d$-dimensions. The renormalization constant $Z_{1}^{1/2}$ corresponds to the
renormalization of the field $\phi$ and is known at five loops in the more
general theory with an internal $O(N)$ symmetry \cite{kle}. The lower order
calculations were performed in \cite{kub}. In our notation defining
\begin{equation}
\gamma(g) ~=~ 2\gamma_{\phi}(g) ~=~ M^2 \frac{\partial~~}{\partial M^2} \ln Z_1
\end{equation}
then, for $N$ $=$ $1$,
\begin{equation}
\gamma(g) ~=~ \frac{1}{6}g^2 ~-~ \frac{1}{8}g^3 ~+~ \frac{65}{81}g^4
\end{equation}
The remaining renormalization constant $Z_{\cal O}$ defines the renormalization
group function $\gamma_{\cal O}(g)$ via
\begin{equation}
\gamma_{\cal O}(g) ~=~ M^2 \frac{\partial~~}{\partial M^2} \ln Z_{\cal O}
\label{ZO}
\end{equation}
which corresponds to the renormalization of the operator insertion in the one
particle irreducible $2$-point Green's function. From the nature of the quartic
interaction $\gamma_{\cal O}(g)$ will have the $O(g^2)$ term as the first
non-zero contribution. In the analogous calculation in QCD we note that the
twist-$2$ operators are physical operators and therefore their anomalous
dimensions are gauge independent. However the renormalization group functions
corresponding to $\gamma(g)$ and $\gamma_{\cal O}(g)$ are each gauge dependent.
Therefore although there is no gauge symmetry in $\phi^4$ it is $\gamma_n(g)$
which we will regard as being on the same footing as the gauge independent
dimensions in QCD.

We now detail the computation of the two and three loop contributions to
(\ref{ZO}). These results will play an important role in the four loop
calculation where they will enter multiplied by vertex counterterms for
example. Although strictly it is their $\epsilon$-expansion to $O(\epsilon)$
and $O(1)$ respectively for two and three loops which we will need it is in
fact possible to compute the relevant integrals exactly as a function of $d$.
The two loop contribution to (\ref{ZO}) arises from the first graph of fig. 1
where the line with a dot on it represents the insertion of the operator
$T_n(p,\Delta)$. There is a flow of momenta $p$ through the external legs of
the Green's function. As we are interested only in the ultra-violet
divergence of the graphs we compute with massless propagators which in turn
allows us to exploit massless integration techniques such as uniqueness
\cite{vas,kaz}. However for the lower order cases one needs only to apply, for
instance, the elementary chain rule of \cite{vas}. Therefore the exact result
for the {\em value} the $2$-loop graph of fig. 1 is
\begin{equation}
\nu(1,1,2\mu-2) \nu_{0nn}(2-\mu,2,3\mu-4+n)
\label{twoloop}
\end{equation}
where $d$ $=$ $2\mu$ $=$ $4$ $-$ $2\epsilon$,
\begin{equation}
\nu(\alpha,\beta, \gamma) ~=~ \pi^\mu a(\alpha) a(\beta) a(\gamma) ~~,~~
\nu_{mnp}(\alpha,\beta, \gamma) ~=~ \pi^\mu a_m(\alpha) a_n(\beta) a_p(\gamma)
\label{nudefns}
\end{equation}
$a(\alpha)$ $=$ $\Gamma(\mu-\alpha)/\Gamma(\alpha)$ and
$a_n(\alpha)$ $=$ $\Gamma(\mu-\alpha+n)/\Gamma(\alpha)$. The $p$ and $\Delta$
dependence is easy to reconstruct from the integral dimension. Performing
the $\epsilon$-expansion of (\ref{twoloop}) and including the symmetry
factor of $1/2$ gives the contribution to (\ref{ZO}) as
\begin{equation}
\frac{g^2}{2n(n+1)\epsilon}
\end{equation}

The three loop calculation requires the calculation of the second graph of fig.
1. As there are two graphs with the same topology we have included both in the
same diagram with their respective names, $E1$ and $E2$. Again it is
straightforward to compute each integral exactly using the chain rule of
\cite{vas} and we find that their values are
\begin{eqnarray}
E1 &=& (\nu(1,1,2\mu-2))^2\nu_{n0n}(2,4-2\mu,4\mu-6+n)
\\
E2 &=& \nu(1,1,2\mu-2)\nu_{n0n}(2,1,2\mu-3+n)\nu_{n0n}(5-2\mu,1,4\mu-6+n)
\end{eqnarray}
However each graph has at least one subgraph divergence arising when either
one or both loops can be enclosed by a box. The divergence arising from these
subgraphs are cancelled by multiplying the two loop graphs of fig. 1 by the
relevant (vertex) counterterm. In this and the four loop case we include in
our expressions for the final value of the divergent part of the graphs the
subtraction of these subgraph divergences. (Further background to this
procedure is well documented in, for example, \cite{lesh}.) However it is
appropriate to record that for $E2$ the subtraction is
\begin{equation}
a^2(1)a_n(2)a_n(3\mu-4+n)/\epsilon
\end{equation}
Hence the full $\MSbar$ contributions to (\ref{ZO}) at $O(g^3)$ are
\begin{eqnarray}
E1 &=& -~ \frac{1}{3n(n+1)\epsilon^2} ~+~ \frac{(2n^2-1)}{3n^2(n+1)^2\epsilon}
\\
E2 &=& -~ \frac{1}{6n(n+1)\epsilon^2} ~+~ \left( - \, \frac{S_1(n)}{3n(n+1)}
+\frac{(4n^2+2n-1)}{6n^2(n+1)^2} \right) \frac{1}{\epsilon}
\end{eqnarray}
where the finite sums $S_l(n)$ are defined as $S_l(n)$ $=$
$\sum_{j=1}^n 1/j^l$. With these values we deduce that
\begin{equation}
\gamma_{\cal O}(g) ~=~ \frac{1}{n(n+1)}g^2 ~+~
\left( \frac{2S_1(n)}{n(n+1)} ~-~ \frac{(10n^2+4n-3)}{2n^2(n+1)^2} \right) g^3
\label{3loop}
\end{equation}

\sect{Four loop calculation.}

We now turn to the renormalization of the operator at the four loop level.
There are four basic topologies to be considered which are illustrated in
fig. 2. As in fig. 1 we have indicated the location of the operator insertion
by a dot and denoted the corresponding graph by the symbol beside each
insertion. Of these four topologies sets $C$ and $D$ represent integrals whose
divergence structure and subgraph subtractions can be computed directly by
elementary chain integrals involving (\ref{nudefns}) and we will not discuss
them further. Likewise integral $A2$ is trivial to compute. The remaining
graphs required some ingenuity and we detail the techniques for several as an
aid to the interested reader and a potential method for future similar
problems.

First, we consider the integral $A1$ and detail its calculation. It can quickly
be reduced to the $2$-loop integral illustrated in the first graph of fig. 3
where the factor arising from the integrations is 
\begin{equation} 
\frac{a_n(1+3\epsilon)(a(1))^3}{4\epsilon^2(1-2\epsilon)(n+1-5\epsilon) 
a(1+\epsilon)a_n(1+4\epsilon)} 
\label{a1fact} 
\end{equation} 
In fig. 3 we denote the power of the propagator beside each line if it is not
unity and the location of the numerator $(\Delta y)^n$ by the symbol $(n)$ and
an arrow on the line beside the propagator involving $y$. We have used the
coordinate space representation of the integrals so that the variables of
integration are the location of the vertices and {\em not} the flow of momentum
around a loop, \cite{vas}. On dimensional grounds the value of this integral is
\begin{equation} 
A_1(\epsilon,n) \frac{(\Delta p)^n}{p^2} \left( \frac{M^2}{p^2}
\right)^{3\epsilon}
\end{equation} 
and the aim is to deduce $A_1(\epsilon,n)$ to the term linear in $\epsilon$.
This is because (\ref{a1fact}) is $O(1/\epsilon^2)$. The first step is to 
relate $A_1(\epsilon,n)$ to an integral with all exponents unity or 
$O(\epsilon)$ which is achieved by the integration by parts rule given in 
\cite{vas,cht}. The result is the set of integrals on the right side of the 
first equation of fig. 3. The three terms with a line missing are easy to 
compute exactly and give the contribution to $A_1(\epsilon,n)$ of
\begin{equation}
\frac{a(1)a(1+\epsilon)a_n(1+2\epsilon)}{6\epsilon(n+1-4\epsilon)
a_n(1+3\epsilon)} \left[ \frac{a_n(1)}{(n-\epsilon)a_n(1+2\epsilon)} ~-~
\frac{a(1)}{(1-3\epsilon)a(1+2\epsilon)} \right]
\end{equation}
The remaining two diagrams have similar structure. If we denote by
$b_1(\epsilon,n)$ the value of the first graph on the right side of fig. 3 then
we can obtain a recurrence relation for $b_1(\epsilon,n)$ with respect to $n$
which can be solved. One way of achieving this is to consider the integral
without the $\Delta$ contraction and decompose the numerator structure
$y_{\mu_1} \ldots y_{\mu_n}$ into a sum of symmetric and traceless objects
$y_{(\mu_1 \ldots \mu_n)}$ defined in \cite{kaz}. As such objects are
independent the resulting integral must be a sum of the set
$y_{(\mu_1 \ldots \mu_r)}$ with $r$ $\leq$ $n$ where the balance of indices is
made up by including the appropriate number of $\eta_{\mu\nu}$ tensors. Their
coefficients are functions of $n$ with the leading term corresponding to
$b_1(\epsilon,n)$. This is deduced by eliminating the next to leading order
term between the equations obtained by first contracting with $\Delta_{\mu_1}
\ldots \Delta_{\mu_n}$ and applying the differential operator
$x_\mu\partial/\partial \Delta^\mu$ and second by contracting with, say,
$\eta^{{\mu_{n-1}}{\mu_n}}$. This results in the relation
\begin{equation}
b_1(\epsilon,n) ~=~ r_n [ b_1(\epsilon,n-1) ~+~ c_1(\epsilon,n-1) ]
\label{a1rr}
\end{equation}
where $c_1(\epsilon,n)$ is defined in the second equation of fig. 3 and
\begin{equation}
r_n ~=~ \frac{(n+\mu-2)}{(n+2\mu-3)}
\end{equation}
Clearly the solution to (\ref{a1rr}) is
\begin{equation}
b_1(\epsilon,n) ~=~ b_1(\epsilon,0) \left( \prod_{i=1}^n r_i \right)
{} ~+~ \sum_{k=1}^n c_1(\epsilon,k-1) \left( \prod_{i=k}^n r_i \right)
\end{equation}
The integral $b_1(\epsilon,0)$ has been computed explicitly in \cite{kazint}
and has the value
\begin{equation}
b_1(\epsilon,0) ~=~ \frac{(1-5\epsilon) a^2(1)a(1+\epsilon)}
{3\epsilon^2(1-3\epsilon)^2(1-4\epsilon) a(1+3\epsilon) }
\end{equation}
in our notation. For $c_1(\epsilon,n)$ the second term is
\begin{equation}
\frac{1}{3(n-\epsilon)(1-3\epsilon)(n+1-3\epsilon)(n+1-4\epsilon)}
\frac{a(1) a(1+\epsilon) a_n(1)}{a_n(1+3\epsilon)}.
\end{equation}
whilst the first and third are similar in form where, for example, it is simple
to derive the integral representation for the first graph of the second
equation of fig. 3 as
\begin{equation}
\frac{1}{3(1-2\epsilon)(1-3\epsilon)}
\frac{\Gamma(1+3\epsilon)}{\Gamma(1+\epsilon)}
\int_0^1 ds \, s^{-\epsilon}(1-s)^{1-2\epsilon}
\int_0^1 dt \, t^{1-3\epsilon}(1-t)^{-\epsilon} [1-st]^{n}
\end{equation}
Since we are only interested in the divergent part of $A1$ it is
straightforward to expand the integral in powers of $\epsilon$ to the
$O(\epsilon)$ term and obtain for it alone
\begin{eqnarray}
&& \int_0^1 {\rm d}s\ s^{-\epsilon}(1-s)^{1-2\epsilon}
\int_0^1 {\rm d}t\ t^{1-3\epsilon}(1-t)^{-\epsilon} [1-st]^{n} \nonumber \\
&& =~ \frac{1}{(n+1)} ~-~ \frac{S_1(n+2)}{(n+1)(n+2)} \nonumber \\
&& +~ \frac{\epsilon}{(n+1)}
\left[3 + S_1(n+2) + \frac{3 S_1(n+2)}{(n+2)} -
\frac{2 S_1^2(n+2)}{(n+2)} - \frac{5 S_2(n+2)}{(n+2)} \right] + O(\epsilon^2)
\label{exparrow}
\end{eqnarray}
Hence we have
\begin{eqnarray}
b_1(\epsilon,n) &=& \frac{(1-2\epsilon)(1-5\epsilon)}
{3\epsilon^2(1-3\epsilon)^2(1-4\epsilon)(n+1-2\epsilon)}
\frac{a_n(1)a^2(1+\epsilon)a(1)}{a(1+3\epsilon)a_n(1+\epsilon)} \nonumber \\
&&+~ \frac{1}{3(n+1)} \left[ 2 - S_2(n+1) - \frac{S_1(n+1)}{(n+1)} \right]
\label{bn}
\end{eqnarray}
Assemblying all the relevant contributions and factors like (\ref{a1fact}) and 
expanding in powers of $\epsilon$ we deduce that the divergent part of $A1$ 
itself is 
\begin{eqnarray}
&& \frac{1}{8n(n+1)\epsilon^3}
{}~+~ \frac{1}{\epsilon^2} \left( \frac{S_1(n)}{4n(n+1)} ~+~
\frac{(5n^2+11n+1)}{8n^2(n+1)^2} \right) \nonumber \\
&&+~ \frac{1}{\epsilon} \left( \frac{(2S_2(n) + S^2_1(n))}{4n(n+1)}
{}~+~ \frac{(5n^2+11n+1)S_1(n)}{4n^2(n+1)^2} \right. \nonumber \\
&& \left. ~~~~~~~+~ \frac{(19n^4+68n^3+85n^2+12n+1)}{8n^3(n+1)^3} \right) ~+~
O(1)
\end{eqnarray}
Performing the simple operation of subtraction of subgraph divergences yields
the result for $A1$ which we have recorded in eqn (\ref{appa1}).

Another technique was also used to compute several of the other graphs of fig.
2 and we detail this for graph $B1$. After performing the two elementary loop 
integrations we obtain a simple two loop graph. Rather than considering this 
graph we examine the more general graph illustrated in fig. 4 where we have 
introduced an extra parameter $a$ and the (analytic) regularization $\delta$. 
Denoting the value of this graph by $B_1(\epsilon,a,\delta)$ we record that for
the problem in hand we require $B_1(\epsilon,1,0)$. The presence of a general 
$a$ and $\delta$ will allow us more freedom in using various calculational 
techniques such as uniqueness to obtain expressions to determine the 
coefficients in the power series.

The first step is to apply the operator $\Delta_\mu \partial/\partial p^\mu$
to the diagram to yield the two graphs $I_1$ and $I_2$ of fig. 4 where
\begin{equation}
B_1(\epsilon,a,\delta) ~=~ \frac{1}{(3+a)\epsilon} [ a \epsilon I_1 ~+~
(1-\delta) I_2 ]
\label{b1defn}
\end{equation}
Whilst the former is finite with respect to $\epsilon$ the second is
divergent. It is determined by integrating by parts on the integral with the
same exponent structure on each line except that $1$ is subtracted from the
top right line and added to the bottom right. Substituting the result of this
operation into (\ref{b1defn}) we obtain
\begin{equation}
B_1(\epsilon,a,\delta) ~=~ \frac{1}{(3+a)\epsilon}
[ (1-\delta)J_2(\epsilon,a,\delta) ~+~ F(\epsilon,a,\delta) ]
\label{b1eqn}
\end{equation}
where $J_2$ is integral given in fig. 4 and $F(\epsilon,a,\delta)$ represents
the sum of the remaining graphs which are finite with respect to $\epsilon$. As
$J_2(\epsilon,a,\delta)$ is a set of elementary chain integrals all that
remains is the calculation of $F(\epsilon,1,0)$ whose $O(\epsilon)$ term is
needed due to the pole in $\epsilon$ in (\ref{b1eqn}). This is achieved by
expanding in powers of $a$ and $\epsilon$
\begin{equation}
F(\epsilon,a,\delta) ~=~ f_0 ~+~ (f_1 + af_2) \epsilon ~+~ O(\epsilon^2)
\end{equation}
and then determining the constant coefficients $f_i$. Rewriting
$F(\epsilon,a,\delta)$ from (\ref{b1eqn}) as
\begin{equation}
F(\epsilon,a,\delta) ~=~ (3+a)\epsilon B_1(\epsilon,a,\delta) ~-~
(1-\delta) J_2(\epsilon,a,\delta) ~ \equiv ~ D(\epsilon,a,\delta)
\end{equation}
then it can be evaluated for special cases of $a$ to give
\begin{equation}
F(\epsilon,1,0) ~=~ 2D(\epsilon,0,0) ~-~ D(\epsilon,-1,0)
\label{Fdefn}
\end{equation}
In the explicit evaluation of the integrals, however, there is a potential
problem when using uniqueness to integrate the right triangle of
$B_1(\epsilon,-1,0)$. To avoid the appearance of factors such as $\Gamma(0)$ in
both numerator and denominator one needs the temporary regulator $\delta$ and
so (\ref{Fdefn}) ought to be replaced by
\begin{equation}
F(\epsilon,1,0) ~=~ [2D(\epsilon,0,\delta) ~-~ D(\epsilon,-1,\delta)]_{\delta
\rightarrow 0}
\end{equation}
As an aid we record the results of these intermediate steps as
\begin{eqnarray}
J_{2}(\epsilon,1,0) &=& \frac{a(1)a(1+\epsilon)a_{n}(1+\epsilon)a_{1}(2)}
{a(2\epsilon)a_{n+1}(1+2\epsilon)} \nonumber \\
D(\epsilon,0,0) &=& \frac{a^3(1)}{a(\epsilon)} \left(
\frac{a_{n}(1+2\epsilon)}{a_{n+1}(1+3\epsilon)} ~-~
\frac{a_{n}(1+\epsilon)}{a_{n+1}(1+2\epsilon)} \right) \nonumber \\
D(\epsilon,-1,0) &=& \frac{a_{n}(1+\epsilon)}
{(1-\epsilon)(n+1-3\epsilon)^{2}a_{n}(1+2\epsilon)}
\end{eqnarray}
With these expressions it is a simple exercise to verify that the result
(\ref{appb1}) is finally obtained for graph $B1$.

We have used either of these algebraic techniques outlined here to evaluate the
remaining divergent parts of the integrals. For completeness the results for
each are recorded in the appendix where the subgraph divergences have been
removed. In several of the results the additional finite sum $K_l(n)$ $=$
$\sum_{j=1}^n (-1)^{(j+1)}/j^l$ occurs. In (\ref{appb3}), for instance, it
arises in the computation of an intermediate two loop integral where the
exponents on the lines diagonally opposite each other are $2$ and $1$ $-$
$\epsilon$ respectively and the central line has zero exponent but an $(n)$
insertion. However, its $\epsilon$-expansion is elementary to compute using the
rule for chains after the application of the binomial expansion. Also in the
graph $A3$ after completing the first two elementary loop integrals one is left
with a two loop integral multiplied by $1/\epsilon$. This integral, which is
finite with respect to $\epsilon$, has been evaluated explicitly in
\cite{kazkot} and involves $K_2(n)$.

Finally it is a simple exercise to assemble all the contributions to the four
loop result and allowing for the symmetry factors of the graphs which are given
in table 1, we find the $O(g^4)$ term of (\ref{3loop}) is
\begin{eqnarray}
&& \frac{3S_2(n)}{n(n+1)} ~+~ \frac{S^2_1(n)}{n^2(n+1)^2}
{}~-~ \frac{(89n^2+53n-18)S_1(n)}{6n^2(n+1)^2} \nonumber \\
&&-~ \frac{(n^2+n-4)K_2(n)}{n^2(n+1)^2}
{}~+~ \frac{(265n^4+280n^3-36n^2-39n+33)}{12n^3(n+1)^3}
\end{eqnarray}
We summarize the results of this and the previous section by recording our
$4$-loop $\MSbar$ value for $\gamma_n(g)$ as
\begin{eqnarray}
\gamma_n(g) &=& -~ \frac{(n-2)(n+3)}{6n(n+1)}g^2 ~+~
\left( \frac{2S_1(n)}{n(n+1)} ~+~ \frac{(n^4+2n^3-39n^2-16n+12)}{8n^2(n+1)^2}
\right)g^3 \nonumber \\
&& +~ \left( \frac{3S_2(n)}{n(n+1)} ~+~ \frac{2S^2_1(n)}{n(n+1)}
{}~-~ \frac{(89n^2+53n-18)S_1(n)}{6n^2(n+1)^2} \right. \nonumber \\
&&-~ \left. \frac{(n^2+n-4)K_2(n)}{n^2(n+1)^2}
{}~+~ \frac{(265n^4+280n^3-36n^2-39n+33)}{12n^3(n+1)^3} ~-~
\frac{65}{81} \! \right) \! g^4
\end{eqnarray}
One non-trivial check on this result is that it vanishes for $n$ $=$ $2$. In
that case the original operator corresponds to the energy momentum tensor
which is conserved in the quantum theory. It is well established that the
anomalous dimension of non-anomalous conserved physical currents vanish to
all orders in perturbation theory. (See, for example, \cite{jcc}.)

\sect{Mellin transform.}

We now apply the results of the previous sections to the problem outlined in
the introduction. Again we emphasise that the aim of the exercise is to
investigate the mathematics underlying the fitting to the Mellin transform of
the anomalous dimensions given only knowledge of the first few moments. In QCD
it is this transform which corresponds to the DGLAP splitting functions,
\cite{dglap}. These are a measure of the probability that a parton decomposes
into other partons ie quarks and gluons. The variable conjugate to the moment
$n$ is Bjorken $x$ which represents the momentum fraction carried by the parton
in the nucleon. Therefore it is restricted to lie in the unit interval. Though
in general the domain of $x$ is the half-line. To approximate the three loop
splitting functions in QCD the authors of \cite{sdis} took a set of trial
functions of $x$ whose properties were consistent with expected general
behaviour of the splitting functions in the unit interval and fixed the unknown
coefficients using the available exact $3$-loop moments. For example, for the
non-singlet case these expressions are known for $n$ $=$ $2$, $4$, $6$, $8$ and
$10$. Of course such an exercise is unnecessary if the full $n$-dependent
result was available. In \cite{sdis} as a prelude to exploring the third order
correction the two loop result was examined. That had the advantage of
knowledge of the exact result of the operator dimension for all $n$ and hence
the splitting functions for all $x$. One disadvantage is that the second order
result has a simpler structure of functions of $x$ than those which would arise
at next order. Also in this second order case the result is unlikely to be
unambiguous as only a few moments are sufficient to isolate which of the set of
trial functions play the important role in, say, the small $x$ behaviour.

Having summarized the status of the QCD calculation we can use our results at
four loops in $\phi^4$ to try and gain a more detailed insight into
improvements to the fitting. First, some elementary remarks are in order. It is
important to note that we are not claiming there is a parton interpretation in
$\phi^4$ theory. The lack of asymptotic freedom counts against this. Second,
although the QCD calculations are at third order and have computed
$\gamma_n(g)$ to four loops, it is in effect a third order calculation since
there is no one loop contribution. Therefore it is this fourth order result
which we will regard as being on the same footing as the three loop QCD result
although we are not claiming that it has as complex a form. We begin by
defining the Mellin transform of a function $f(x)$ as
\begin{equation}
{\cal M}[f(x)] ~=~ \int_0^1 \, dx \, x^{n-1} f(x)
\label{Meldefn}
\end{equation}
where we restrict the function to exist on the unit interval. In QCD the
operator dimension is gauge independent though the field dimension and the
renormalization of the bare operator are gauge dependent. As there is no
gauge symmetry in $\phi^4$ theory rather than take the full dimension
$\gamma_n(g)$ we concentrate on the part $\gamma_{\cal O}(g)$. This is
justified by the fact that $\gamma(g)$ is $n$-independent and therefore its
contribution to the $x$-dependence will be purely $\gamma(g)\delta(1-x)$
which is uninteresting for our study. With the definition (\ref{Meldefn})
it is straightforward to verify that the splitting functions are given by
\begin{equation}
P(g,x) ~=~ \sum_{n=2}^\infty P_n(x) g^n
\end{equation}
where
\begin{eqnarray}
P_2(x) &=& (1-x) \\
P_3(x) &=& -~ 2x \ln(x) ~-~ 2(1-x)\ln(1-x) ~-~ \half (1-x)( 3\ln(x) + 10 ) \\
P_4(x) &=& 8(1-x)\mbox{Li}_3(-x) ~-~ 4(1-x)\ln(x)\mbox{Li}_2(-x) ~-~
3(1-x)\psi^{\prime\prime}(1) \nonumber \\
&& -~ 4x\ln(x)\psi^\prime(1) ~+~ \frac{2}{3}x\ln^3(x) ~+~ 9(1+x)\ln(x)\ln(1+x)
\nonumber \\
&& +~ 9(1+x)\mbox{Li}_2(-x) ~-~ 2(1-x)\mbox{Li}_2(x) ~-~ 4x\mbox{Li}_2(x)
\nonumber \\
&& +~ (1-x)\ln(x)\ln(1-x) ~+~ 2(1-x)\ln^2(1-x) ~+~ \frac{89}{6}(1-x)\ln(1-x)
\nonumber \\
&& +~ 13x\psi^\prime(1) ~+~ 3(1-x)\psi^\prime(1) ~+~ \frac{93}{4}(1-x)
\nonumber \\
&& +~ \frac{23}{2}\ln(x) ~+~ \frac{11}{8}\ln^2(x) ~-~ \frac{27}{8}x\ln^2(x) ~+~
\frac{9}{2}x\ln(x)
\end{eqnarray}
In these expressions $\mbox{Li}_n(x)$ is the polylogarithm function whose
properties are well known, \cite{lew}, and $\psi(x)$ is the logarithmic
derivative of the Euler $\Gamma$-function. To gain some idea into the form of
these functions we have plotted them in figs 5-7.

As a preliminary to the four loop analysis we first summarize the method for
the three loop case since the two loop example is clearly trivial. The basic
idea is to postulate a basis set of trial functions of $x$ with arbitrary
coefficients. These are fixed by using (\ref{Meldefn}) to obtain a function of
$n$ which can be evaluated for the first few moments and compared with the
analogous value of (\ref{3loop}). Once the coefficients have been fixed one can
compare the approximate function with the exact result to see how well it
covers the function in the unit interval. It is worth pointing out that the
trial set of functions ought to contain a few which have logarithmic dependence
given the nature of the exact result and its form in fig. 6. Therefore some
guidance on how good the fit is, is that it should be better than taking a
trial set which gives a polyniomial in $x$ of the appropriate degree. Given
this criterion we discovered that the set
\begin{equation}
\{ \, 1, x, \ln(x), \ln(1-x), x\ln(1-x) \, \}
\label{trial3}
\end{equation}
gave a very good approximation. We have plotted this approximation with the
exact result in fig. 8 where the straight line is the exact result and the dots
represent the approximation. Interestingly the discrepancy at $x$ $=$ $0$ is
about $0.12$. Clearly to achieve this accuracy we have used only five moments.
However, comparing the functions of (\ref{trial3}) with the exact function one
observes that $x\ln x$ is absent. This illustrates a potential pitfall in
fitting approximations to exact results. One feature of the three loop case
to emerge was the necessity of having functions in the trial set which
reproduced some of the functions of $x$ which appeared in the exact result.
We repeated this exercise with a variety of sets either with fewer functions
and therefore fewer moments or replaced several of the entries in
(\ref{trial3}) with other functions. However the result of fig. 8 represents
the best fit.

We have repeated this analysis for the $4$-loop case. From the form of the
exact result it turns out to be a harder exercise especially in covering the
small $x$ region well. Again with the criterion that a fit must be better than
a polynomial, we managed to gain reasonable approximations with two trial sets.
To achieve this we needed to extend the set to seven elements and include
higher powers of the logarithm. Even with this, however, an answer as accurate
as three loops could not be achieved. These sets were
\begin{equation}
\{ \, 1, x, \ln(x), \ln^2(x), \mbox{Li}_2(x), \ln(x)\mbox{Li}_2(x), x\ln(x)
\, \}
\label{trial41}
\end{equation}
\begin{equation}
\{ \, 1, x, \ln(x), \ln^2(x), \mbox{Li}_2(x), \ln(x)\mbox{Li}_2(x),
\mbox{Li}_2(-x) \, \}
\label{trial42}
\end{equation}
and we have plotted each beside the exact results respectively in figs 9 and
10. Although set (\ref{trial42}) appears to deviate in the middle $x$ range
it is clearly more accurate for small $x$ than (\ref{trial41}). This is due
to the alternating form of the expansion of the dilogarithm for small $x$. The
respective discrepancies at $x$ $=$ $0$ from the exact and approximate answers
were $2.5$ and $0.5$.

\sect{Conclusions.}

We have constructed the $\MSbar$ $4$-loop anomalous dimensions for a set of
twist-$2$ operators which are analogous to operators which occur in the
operator product expansion used in deep inelastic scattering in QCD. In the
course of examining the inverse Mellin transform for the four loop case we
found that it was important to isolate those functions of $x$ and include them
in the fit that reproduced similar forms to the basic finite sums that appear
in the exact answer. In the absence of a full exact answer for QCD it would
therefore seem to us that one could endeavour to improve the fits of the
anomalous dimensions, \cite{sdis}, in the low $x$-region by trying to deduce
the type of finite sums which would occur at three loops. An insight into this
could be obtained by generalizing some of the series that already occur in the
exact two loop results. For instance, the sums $S_l(n)$ occur at $l$-loops for
$l$ $=$ $1$ and $2$ and it would appear odd if $S_3(n)$ did not occur in the
$3$-loop result. Therefore another structure to study might arise from
generalizing those finite sums which only occur in the two loop result. The
other feature which we observed in our analysis was that to maintain the
accuracy one would require several more exact moments for the anomalous
dimensions.

\vspace{2cm}
\noindent
{\bf Acknowledgements.} The authors thank the organisers of the conference
`Renormalization Group 96' where this work was initiated. JAG is supported by
PPARC through an Advanced Fellowship, ANM is supported by Grant 97-01-01152 of 
the Russian Fond for Fundamental Research and SED~--~by INTAS Grant 
93--2492--ext. Part of this work was carried out with the aid of the computer 
algebra packages {\sc Reduce}, \cite{red}, and {\sc Maple} versions $V.4$.

\newpage

\appendix

\sect{Values for $4$-loop integrals.}
In this section we record the values of the divergent parts of the integrals
which arise in the computation of the $4$-loop anomalous dimension in $\MSbar$.
The label refers to the diagrams of fig. 2 though the values given here
correspond to those with the subgraph divergences removed. We found

\begin{eqnarray}
A1 &=& \frac{1}{24n(n+1)\epsilon^3}
{}~+~ \left( \frac{S_1(n)}{12n(n+1)} - \frac{(7n^2+5n-1)}{24n^2(n+1)^2}
\right) \frac{1}{\epsilon^2} \nonumber \\
&& +~ \left( -~ \frac{S_2(n)}{6n(n+1)} + \frac{S_1^2(n)}{12n(n+1)}
- \frac{(7n^2+5n-1)S_1(n)}{12n^2(n+1)^2} \right. \nonumber \\
&& \left. ~~~~~ +~ \frac{(25n^4+36n^3+7n^2-4n+1)}{24n^3(n+1)^3} \right)
\frac{1}{\epsilon} \label{appa1} \\
A2 &=& \frac{1}{12n(n+1)\epsilon^3}
{} ~-~ \frac{(5n^2+3n-1)}{12n^2(n+1)^2\epsilon^2}
{} ~+~ \frac{(9n^4+8n^3-3n^2-2n+1)}{12n^3(n+1)^3\epsilon} \\
A3 &=& \frac{K_2(n)}{n^2(n+1)^2\epsilon}
\end{eqnarray}

\begin{eqnarray}
B1 &=& \frac{1}{24n(n+1)\epsilon^3} ~+~ \left( \frac{S_1(n)}{12n(n+1)}
- \frac{(7n^2+5n-1)}{24n^2(n+1)^2} \right) \frac{1}{\epsilon^2} \nonumber \\
&& ~+~ \left( \frac{S_2(n)}{3n(n+1)} + \frac{S_1^2(n)}{12n(n+1)}
- \frac{(7n^2+5n-1)S_1(n)}{12n^2(n+1)^2} \right. \nonumber \\
&& \left. ~~~~~~~~~ +~ \frac{(25n^4+36n^3+7n^2-4n+1)}{24n^3(n+1)^3} \right)
\frac{1}{\epsilon} \label{appb1} \\
B2 &=& \frac{1}{8n(n+1)\epsilon^3}
{} ~+~ \frac{1}{\epsilon^2} \left( \frac{S_1(n)}{12n(n+1)}
- \frac{(11n^2+5n-3)}{24n^2(n+1)^2} \right) \nonumber \\
&& ~+~ \left( \frac{S_2(n)}{12n(n+1)} - \frac{S_1^2(n)}{6n(n+1)}
- \frac{S_1(n)(n^2-n-1)}{12n^2(n+1)^2} \right. \nonumber \\
&& \left. ~~~~~~~~~ +~ \frac{(13n^4+4n^3-11n^2-2n+3)}{24n^3(n+1)^3} \right)
\frac{1}{\epsilon} \\
B3 &=& \frac{1}{12n(n+1)\epsilon^3} ~+~ \left( \frac{S_1(n)}{6n(n+1)}
- \frac{(4n^2+2n-1)}{12n^2(n+1)^2} \right) \frac{1}{\epsilon^2} \nonumber \\
&& ~+~ \left( \frac{S_2(n)}{6n(n+1)} + \frac{S_1^2(n)}{6n(n+1)}
- \frac{S_1(n)(4n^2+2n-1)}{6n^2(n+1)^2} \right. \nonumber \\
&& \left. ~~~~~~~~~ -~ \frac{K_2(n)}{2n(n+1)}
+ \frac{(4n^4-5n^2-n+1)}{12n^3(n+1)^3} \right)
\frac{1}{\epsilon} \label{appb3}
\end{eqnarray}

\begin{eqnarray}
C1 &=& \frac{1}{4n(n+1)\epsilon^3} ~-~ \frac{(2n^2-1)}{4n^2(n+1)^2\epsilon^2}
{} ~-~ \frac{(4n^3+3n^2-n-1)}{4n^3(n+1)^3\epsilon} \\
C2 &=& C3 ~=~ \frac{1}{12n(n+1)\epsilon^3} ~+~ \left( \frac{S_1(n)}{6n(n+1)}
- \frac{(4n^2+2n-1)}{12n^2(n+1)^2} \right) \frac{1}{\epsilon^2} \nonumber \\
&& ~+~ \left( \frac{S_2(n)}{6n(n+1)} + \frac{S_1^2(n)}{6n(n+1)}
- \frac{S_1(n)(4n^2+2n-1)}{6n^2(n+1)^2} \right. \nonumber \\
&& \left. ~~~~~~~~~ +~ \frac{(4n^4-5n^2-n+1)}{12n^3(n+1)^3} \right)
\frac{1}{\epsilon}
\end{eqnarray}

\begin{eqnarray}
D1 &=& \frac{1}{16n(n+1)\epsilon^2}
{}~-~ \frac{(13n^2+9n-2)}{32n^2(n+1)^2\epsilon} \\
D2 &=& \frac{1}{16n(n+1)\epsilon^2} ~+~ \left( \frac{S_1(n)}{8n(n+1)}
- \frac{(3n+2)(3n+1)}{32n^2(n+1)^2} \right) \frac{1}{\epsilon} \\
D3 &=& -~ \frac{1}{8n^2(n+1)^2\epsilon^2}
{}~-~ \frac{(n^2-n-1)}{4n^3(n+1)^3\epsilon}
\end{eqnarray}

\newpage

\begin{tabular}{c||c|c|c|c|c|c|c|c|c|c|c|c}
Graph & A1 & A2 & A3 &
B1 & B2 & B3 &
C1 & C2 & C3 &
D1 & D2 & D3 \\
\hline
Factor & 1 & 1/4 & 1/2 &
1 & 1/2 & 1/4 &
1/8 & 1/2 & 1/4 &
1/6 & 1/6 & 1/4 \\
\end{tabular}

\vspace{1.5cm}
{\bf Table 1. Symmetry factors for each graph defined in fig. 2.}

\newpage
\newpage
\noindent
{\Large {\bf Figure Captions.}}
\begin{description}
\item[Fig. 1.] Two and three loop graphs contributing to $\gamma_n(g)$.
\item[Fig. 2.] Four loop graphs contributing to $\gamma_n(g)$.
\item[Fig. 3.] Intermediate steps in the calculation of $A1$.
\item[Fig. 4.] Intermediate integrals in the calculation of $B1$.
\item[Fig. 5.] Two loop splitting function.
\item[Fig. 6.] Three loop splitting function.
\item[Fig. 7.] Four loop splitting function.
\item[Fig. 8.] Approximation to the three loop splitting function.
\item[Fig. 9.] Approximation to the four loop splitting function with first
trial set.
\item[Fig. 10.] Approximation to the four loop splitting function with second
trial set.
\end{description}

\newpage

\unitlength=0.150mm

\begin{picture}(400.00,400.00)
\put(110.00,200.00){\circle{200.00}}
\put(110.00,200.00){\circle*{12.00}}
\put(10.00,200.00){\line(1,0){200}}

\bezier{300}(410.00,200.00)(510.00,400.00)(610.00,200.00)
\bezier{300}(410.00,200.00)(460.00,270.00)(510.00,200.00)
\bezier{300}(410.00,200.00)(460.00,130.00)(510.00,200.00)
\bezier{300}(510.00,200.00)(560.00,270.00)(610.00,200.00)
\bezier{300}(510.00,200.00)(560.00,130.00)(610.00,200.00)
\put(400,200){\line(1,0){10}}
\put(610,200){\line(1,0){10}}
\put(460.00,165.00){\circle*{12}}
\put(510.00,300.00){\circle*{12}}
\put(500.00,320.00){$ E1  $}
\put(450.00,120.00){$ E2  $}
\end{picture}

\noindent
{\bf Fig. 1. Two and three loop graphs contributing to $\gamma_n(g)$.}


\begin{picture}(800.00,450.00)
\put(00.00,50.00){
\begin{picture}(510.00,400.00)
\put(110.00,200.00){\oval(200,200)[t]}
\put(10.00,200.00){\line(3,2){100}}
\put(210.00,200.00){\line(-3,2){100}}
\put(10.00,200.00){\line(3,-2){100}}
\put(210.00,200.00){\line(-3,-2){100}}
\put(60.00,233.33){\circle*{12.00}}
\put(45.00,253.33){$ A1 $}
\put(110.00,300){\circle*{12.00}}
\put(92.00,320){$ A2 $}
\put(80.00,200){\circle*{12.00}}
\put(90.00,190){$ A3 $}
\bezier{300}(110.00,266.66)(50.00,200.00)(110.00,133.33)
\bezier{300}(110.00,266.66)(170.00,200.00)(110.00,133.33)
\put(00.00,200.00){\line(1,0){10}}
\put(210.00,200.00){\line(1,0){10}}

\put(910.00,200.00){\line(-3,2){100}}
\put(810.00,266.66){\line(0,-1){133.33}}
\put(710.00,200.00){\line(3,-2){100}}
\put(750.00,246.33){\circle*{12.00}}
\put(740.00,270.33){$ B1 $}
\put(760.00,166.66){\circle*{12.00}}
\put(740.00,130.66){$ B2 $}
\put(810.00,200.00){\circle*{12.00}}
\put(820.00,200.00){$ B3 $}
\bezier{300}(710.00,200.00)(740.00,263.00)(810.00,266.66)
\bezier{300}(710.00,200.00)(780.00,203.00)(810.00,266.66)
\bezier{300}(910.00,200.00)(880.00,136.00)(810.00,133.33)
\bezier{300}(910.00,200.00)(840.00,199.00)(810.00,133.33)
\put(700.00,200.00){\line(1,0){10}}
\put(910.00,200.00){\line(1,0){10}}

\bezier{300}(10.00,-200.00)(160.00,0.00)(310.00,-200.00)
\bezier{300}(10.00,-200.00)(60.00,-130.00)(110.00,-200.00)
\bezier{300}(10.00,-200.00)(60.00,-270.00)(110.00,-200.00)
\bezier{300}(110.00,-200.00)(160.00,-130.00)(210.00,-200.00)
\bezier{300}(110.00,-200.00)(160.00,-270.00)(210.00,-200.00)
\bezier{300}(210.00,-200.00)(260.00,-130.00)(310.00,-200.00)
\bezier{300}(210.00,-200.00)(260.00,-270.00)(310.00,-200.00)
\put(0,-200){\line(1,0){10}}
\put(310,-200){\line(1,0){10}}
\put(60.00,-235.00){\circle*{12}}
\put(160.00,-235.00){\circle*{12}}
\put(160.00,-100.00){\circle*{12}}
\put(150.00,-80.00){$ C1  $}
\put(45.00,-280.00){$ C2  $}
\put(145.00,-280.00){$ C3  $}

\bezier{300}(610.00,-200.00)(760.00,0.00)(910.00,-200.00)
\bezier{300}(610.00,-200.00)(760.00,-400.00)(910.00,-200.00)
\put(760.00,-200.00){\circle{120}}
\put(600.00,-200.00){\line(1,0){10}}
\put(910.00,-200.00){\line(1,0){10}}
\put(610.00,-200.00){\line(1,0){300}}
\put(660.00,-200.00){\circle*{12}}
\put(760.00,-152.00){\circle*{12}}
\put(760.00,-100.00){\circle*{12}}
\put(750.00,-80.00){$ D1  $}
\put(650.00,-180.00){$ D2  $}
\put(750.00,-135.00){$ D3  $}
\end{picture}}
\end{picture}

\vspace{5cm}
\noindent
{\bf Fig. 2. Four loop graphs contributing to $\gamma_n(g)$.}

\newpage

\unitlength=0.150mm
\begin{picture}(600.00,750.00)
\put(00.00,300.00){
\begin{picture}(210.00,400.00)
\put(10.00,200.00){\line(3,2){100}}
\put(10.00,200.00){\vector(3,2){50}}
\put(210.00,200.00){\line(-3,2){100}}
\put(10.00,200.00){\line(3,-2){100}}
\put(210.00,200.00){\line(-3,-2){100}}
\put(50.00,250.00){${\scriptstyle (n)}$}
\put(60.00,210.00){${\scriptstyle 2}$}
\put(110.00,266.66){\line(0,-1){133.33}}
\put(120.00,200.00){$ \epsilon $}
\end{picture}}
\put(250.00,495.00){$=~-(1-3\epsilon+n)$}

\put(440.00,300.00){
\begin{picture}(210.00,400.00)
\put(10.00,200.00){\line(3,2){100}}
\put(10.00,200.00){\vector(3,2){50}}
\put(210.00,200.00){\line(-3,2){100}}
\put(10.00,200.00){\line(3,-2){100}}
\put(210.00,200.00){\line(-3,-2){100}}
\put(50.00,250.00){${\scriptstyle (n)}$}
\put(110.00,266.66){\line(0,-1){133.33}}
\put(120.00,200.00){$ \epsilon $}
\end{picture}}

\put(690.00,495.00){$+\>n$}

\put(740.00,300.00){
\begin{picture}(210.00,400.00)
\put(110.00,200.00){\oval(200,200)[t]}
\put(100.00,320.00){${\scriptstyle 0;\> (1)} $}
\put(80.00,300){\vector(1,0){30}}
\put(30.00,250.00){${\scriptstyle (n-1)}$}
\put(10.00,200.00){\line(3,2){100}}
\put(10.00,200.00){\vector(3,2){50}}
\put(210.00,200.00){\line(-3,2){100}}
\put(10.00,200.00){\line(3,-2){100}}
\put(210.00,200.00){\line(-3,-2){100}}
\put(110.00,266.66){\line(0,-1){133.33}}
\put(120.00,200.00){$ \epsilon $}
\end{picture}}

\put(150.00,245.00){$+\>$}
\put(190.00,50.00){
\begin{picture}(210.00,400.00)
\put(10.00,200.00){\line(3,2){100}}
\put(10.00,200.00){\vector(3,2){50}}
\put(10.00,200.00){\line(3,-2){100}}
\put(210.00,200.00){\line(-3,-2){100}}
\put(50.00,250.00){${\scriptstyle (n)}$}
\put(60.00,210.00){${\scriptstyle 2}$}
\put(110.00,266.66){\line(0,-1){133.33}}
\put(120.00,200.00){$ \epsilon $}
\end{picture}}
\put(440.00,245.00){$+\>\epsilon\>\> \Biggl ( $}

\put(500.00,50.00){
\begin{picture}(210.00,400.00)
\put(10.00,200.00){\line(3,2){100}}
\put(10.00,200.00){\vector(3,2){50}}
\put(10.00,200.00){\line(3,-2){100}}
\put(210.00,200.00){\line(-3,-2){100}}
\put(50.00,250.00){${\scriptstyle (n)}$}
\put(110.00,266.66){\line(0,-1){133.33}}
\put(120.00,200.00){$ 1+\epsilon $}
\end{picture}}

\put(740.00,245.00){$-$}

\put(760.00,50.00){
\begin{picture}(210.00,400.00)
\put(50.00,250.00){${\scriptstyle (n)}$}
\put(10.00,200.00){\line(3,2){100}}
\put(10.00,200.00){\vector(3,2){50}}
\put(210.00,200.00){\line(-3,2){100}}
\put(10.00,200.00){\line(3,-2){100}}
\put(110.00,266.66){\line(0,-1){133.33}}
\put(120.00,200.00){$1+\epsilon $}
\end{picture}}
\put(980.00,245.00){$\Biggr )$}

\put(-10.00,15.00){$c_1(\epsilon,n) ~=~ $}
\put(95.00,-180.00){
\begin{picture}(210.00,400.00)
\put(10.00,200.00){\line(3,2){100}}
\put(10.00,200.00){\vector(3,2){50}}
\put(210.00,200.00){\line(-3,2){100}}
\put(10.00,200.00){\line(3,-2){100}}
\put(210.00,200.00){\line(-3,-2){100}}
\put(50.00,250.00){${\scriptstyle (n)}$}
\put(60.00,210.00){${\scriptstyle 0}$}
\put(110.00,266.66){\line(0,-1){133.33}}
\put(120.00,200.00){$ \epsilon $}
\end{picture}}
\put(330.00,15.00){ $ - $ }

\put(360.00,-180.00){
\begin{picture}(210.00,400.00)
\put(10.00,200.00){\line(3,2){100}}
\put(10.00,200.00){\vector(3,2){50}}
\put(10.00,200.00){\line(3,-2){100}}
\put(210.00,200.00){\line(-3,-2){100}}
\put(50.00,250.00){${\scriptstyle (n)}$}
\put(110.00,266.66){\line(0,-1){133.33}}
\put(120.00,200.00){$ \epsilon $}
\end{picture}}

\put(600.00,15.00){$~-~ \mbox{\LARGE {$\frac{n}{(n+\mu-1)}$} } $}

\put(760.00,-180.00){
\begin{picture}(210.00,400.00)
\put(28.00,250.00){${\scriptstyle (n-1)}$}
\put(10.00,200.00){\line(3,2){100}}
\put(10.00,200.00){\vector(3,2){50}}
\put(210.00,200.00){\line(-3,2){100}}
\put(60.00,210.00){${\scriptstyle 0}$}
\put(10.00,200.00){\line(3,-2){100}}
\put(210.00,200.00){\line(-3,-2){100}}
\put(110.00,266.66){\line(0,-1){133.33}}
\put(120.00,200.00){$ \epsilon $}
\end{picture}}
\end{picture}

\vspace{3cm}
\noindent
{\bf Fig. 3. Intermediate steps in the calculation of $A1$.}

\newpage

\unitlength=0.150mm

\begin{picture}(600,600)

\put(100,300){
\begin{picture}(210.00,300.00)
\put(10.00,200.00){\line(3,2){100}}
\put(10.00,200.00){\vector(3,2){50}}
\put(210.00,200.00){\line(-3,2){100}}
\put(10.00,200.00){\line(3,-2){100}}
\put(210.00,200.00){\line(-3,-2){100}}
\put(10.00,240.00){${1+\epsilon}$}
\put(50.00,210.00){${\scriptstyle (n)}$}
\put(110.00,266.66){\line(0,-1){133.33}}
\put(160.00,240.00){$1- \delta $}
\put(160.00,150.00){$a\epsilon+\delta $}
\put(-160.00,195.00){$B_1(\epsilon,a,\delta) ~=~$}
\end{picture}}

\put(600,300){
\begin{picture}(210.00,300.00)
\put(10.00,200.00){\line(3,2){100}}
\put(10.00,200.00){\vector(3,2){50}}
\put(110.00,133.33){\vector(3,2){50}}
\put(210.00,200.00){\line(-3,2){100}}
\put(10.00,200.00){\line(3,-2){100}}
\put(210.00,200.00){\line(-3,-2){100}}
\put(10.00,240.00){${1+\epsilon}$}
\put(50.00,210.00){${\scriptstyle (n)}$}
\put(110.00,266.66){\line(0,-1){133.33}}
\put(160.00,240.00){$1- \delta $}
\put(160.00,150.00){$1+a\epsilon+\delta $}
\put(-160.00,195.00){$I_{1}(\epsilon,a,\delta) ~=~$}
\end{picture}}

\put(100,0){
\begin{picture}(210.00,300.00)
\put(10.00,200.00){\line(3,2){100}}
\put(10.00,200.00){\vector(3,2){50}}
\put(110.00,266.66){\vector(3,-2){50}}
\put(210.00,200.00){\line(-3,2){100}}
\put(10.00,200.00){\line(3,-2){100}}
\put(210.00,200.00){\line(-3,-2){100}}
\put(10.00,240.00){${1+\epsilon}$}
\put(50.00,210.00){${\scriptstyle (n)}$}
\put(110.00,266.66){\line(0,-1){133.33}}
\put(160.00,240.00){$2- \delta $}
\put(160.00,150.00){$a\epsilon+\delta $}
\put(-160.00,195.00){$I_{2}(\epsilon,a,\delta) ~=~$}
\end{picture}}

\put(600,100){
\begin{picture}(210.00,300.00)
\put(10.00,100.00){\line(3,2){100}}
\put(10.00,100.00){\vector(3,2){50}}
\put(210.00,100.00){\line(-3,2){100}}
\put(10.00,100.00){\line(3,-2){100}}
\put(210.00,100.00){\line(-3,-2){100}}
\put(70.00,120.00){${\scriptstyle (n)}$}
\put(120.00,120.00){${\scriptstyle (1)}$}
\put(110.00,166.66){\vector(3,-2){50.00}}
\put(15.00,150.00){$1+\epsilon $}
\put(160.00,140.00){$2- \delta $}
\put(160.00,50.00){$1+a\epsilon+\delta $}
\put(-160.00,95.00){$J_{2}(\epsilon,a,\delta) ~=~$}
\end{picture}}

\end{picture}

\vspace{1cm}
\noindent
{\bf Fig. 4. Intermediate integrals in the calculation of $B1$.}

\newpage
\epsfxsize=12cm
\epsfbox{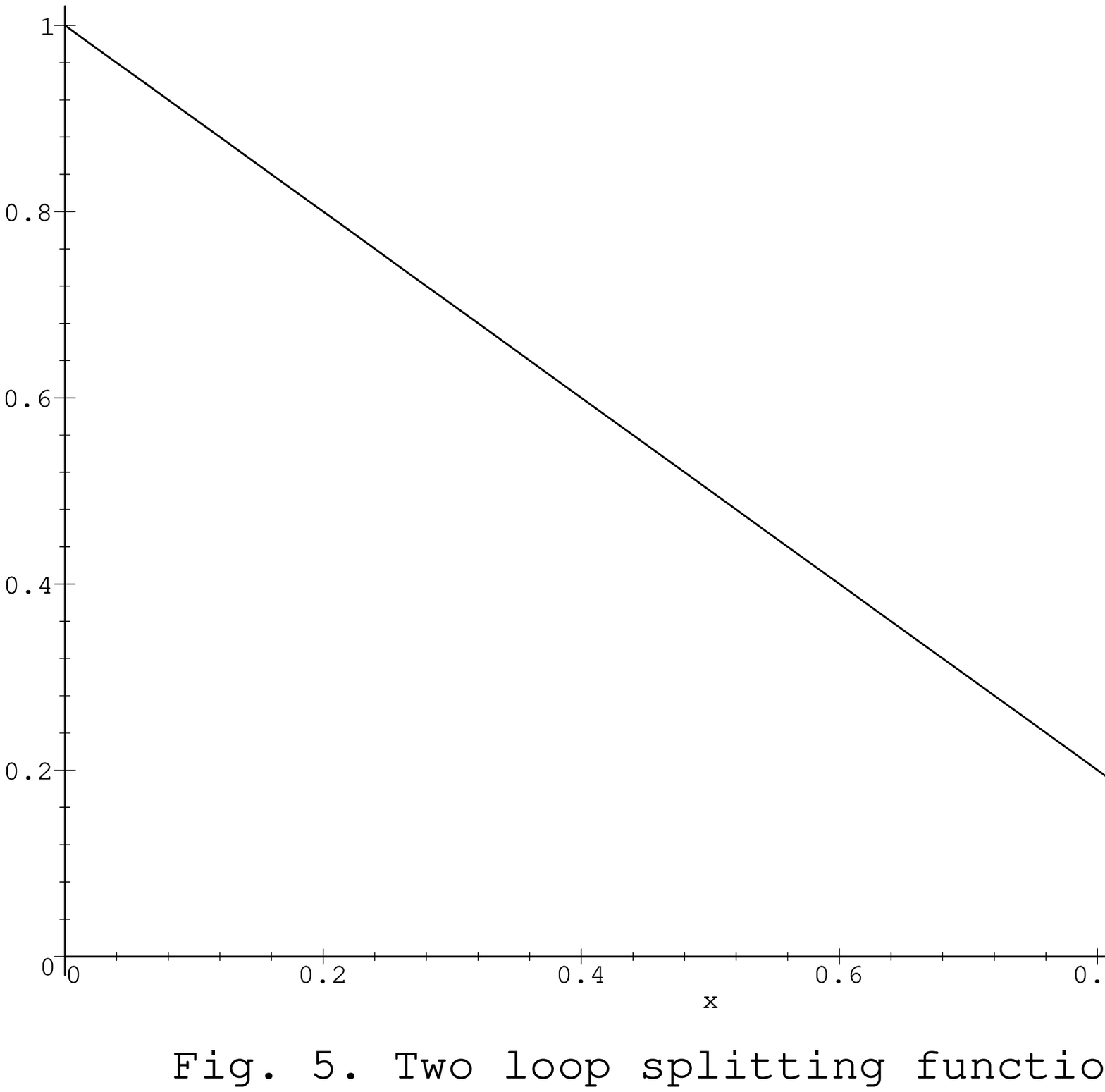}
\pagestyle{empty}

\newpage
\epsfxsize=12cm
\epsfbox{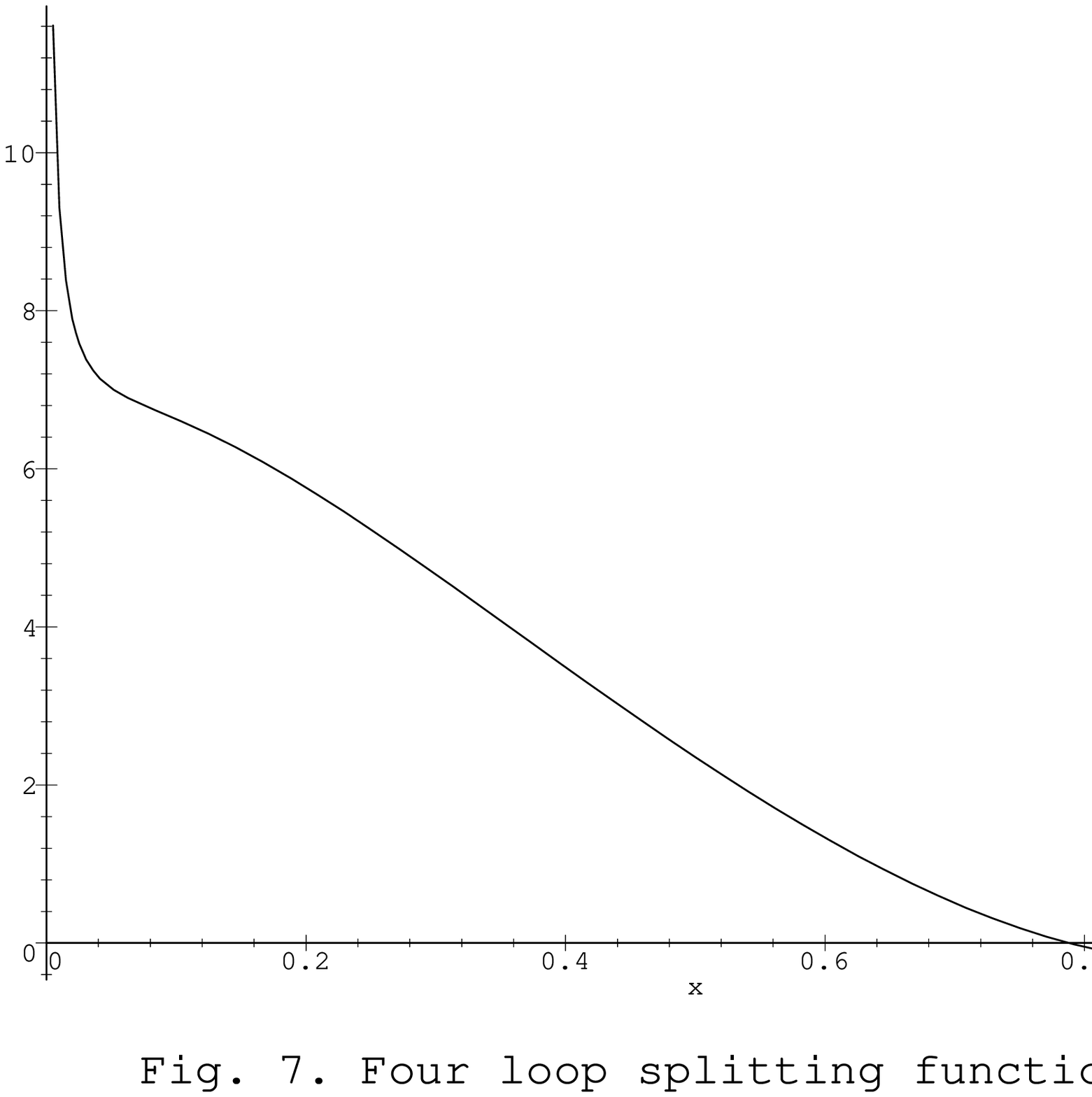}

\newpage
\epsfxsize=12cm
\epsfbox{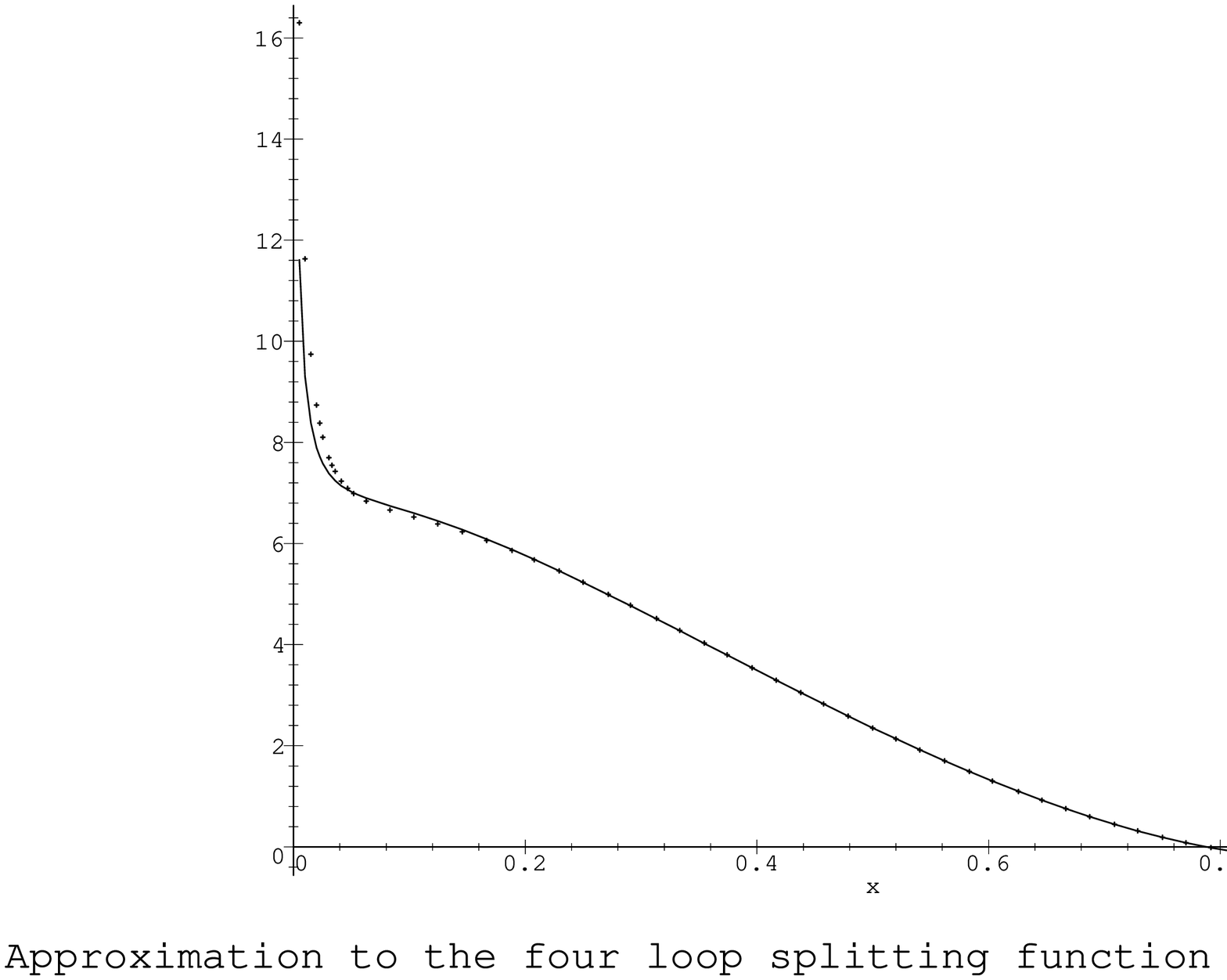}

\end{document}